\documentclass[aps, prb, superscriptaddress, reprint, nobibnotes, twocolumn]{revtex4-2}
\usepackage{amsmath,amssymb}
\usepackage{color}
\usepackage{comment}
\usepackage{bm,graphicx,hyperref}
\hypersetup{%
  breaklinks = {true},
  citecolor = {blue},
  colorlinks = {true},
  linkcolor = {red},}

\usepackage{lineno}

\usepackage{cancel}

\begin{document}

\title{Local-time formula for dissipation in solid ionic electrolytes}

\author{A. Rodin}
\affiliation{Yale-NUS College, 16 College Avenue West, 138527, Singapore}
\affiliation{Centre for Advanced 2D Materials, National University of Singapore, 117546, Singapore}
\affiliation{Department of Materials Science and Engineering, National University of Singapore, 117575, Singapore}

\author{B. A. Olsen}
\affiliation{Department of Physics, Lewis \& Clark College, Portland, Oregon, 97219, USA}

\author{A. Ustyuzhanin}
\affiliation{Constructor University, Bremen, Campus Ring 1, 28759, Germany}
\affiliation{Institute for Functional Intelligent Materials, National University of Singapore, 4 Science Drive 2, Singapore 117544, Singapore}

\author{A. Maevskiy}
\affiliation{Institute for Functional Intelligent Materials, National University of Singapore, 4 Science Drive 2, Singapore 117544, Singapore}

\author{K. Noori}
\affiliation{Institute for Functional Intelligent Materials, National University of Singapore, 4 Science Drive 2, Singapore 117544, Singapore}

\begin{abstract}

When ions move through solids, they interact with the solid's constituent atoms and cause them to vibrate around their equilibrium points.
This vibration, in turn, modifies the potential landscape through which the mobile ions travel.
Because the present-time potential depends on past interactions, the coupling is inherently non-local in time, making its numerical and analytical treatment challenging.
For sufficiently slow-moving ions, we linearize the phonon spectrum to show that these non-local effects can be ignored, giving rise to a drag-like force.
Unlike the more familiar drag coefficient in liquids, the drag takes on a matrix form due to the crystalline structure of the framework. 
We numerically simulate trajectories and dissipation rates using both the time-local and non-local formulas to validate our simplification.
The time-local formula dramatically reduces the computational cost of calculating the motion of mobile particle through a crystalline framework and clearly connects the properties of the material to the drag experienced by the particle.

\end{abstract}	

\maketitle

\section{Introduction}
\label{sec:Introduction}

Understanding ionic motion through solids~\cite{MahanRoth, Mehrer} is essential to developing high-performance solid-state electrolytes necessary for the operation of solid-state batteries~\cite{Bachman2016, Manthiram2017, Famprikis2019}.
A substantial fraction of theoretical work focuses on elucidating the mechanisms behind ionic current flow~\citep{Deng2015, Marcolongo2017, He2017, DiStefano2019, Fall2019, Poletayev2022, Carvalho2022, Hu2023} and how the properties of the solid framework impact the ionic transport, in particular the ability of the current-carrying ions to overcome energy barriers within the host lattice~\citep{Wang2015, Bo2016, Krauskopf2017, Brenner2020}.
A complementary and equally important question is how these ions lose energy to the solid framework during their motion.
Relating this energy loss and, by extension, the resistance to the material properties is crucial for designing application-grade electrolytes.
The fundamental difficulty of predicting material resistivity is that the dynamics of the mobile ion and the framework occur on similar time scales.
As a result, the mobile particle both displaces and experiences recoil from the framework, which is continuously responding to the mobile particle's earlier motion, leading to equations of motion that are not local in time.
In this work, we develop an approximation scheme to circumvent this difficulty, so the dissipation is determined from the framework's unperturbed potential landscape and its vibrational modes.

Earlier work~\citep{Rodin2022} introduced an analytic formula for dissipation in three-dimensional (3D) crystals using a local-time approximation.
The results showed that the drag experienced by the mobile particles depends on the curvature of the potential generated by the lattice.
Additionally, the dissipation rate is inversely proportional to the cube of speed of sound in the material and also to the density of the crystal.
In other words, the dissipation is lowest in stiff, dense materials.
Due to the complexity of the problem, these results were not validated using numerical demonstrations.

Following the analytical work of Ref.~\citep{Rodin2022}, dissipation has been explored in the simplest crystal functioning as a solid electrolyte: an infinite one-dimensional (1D) chain of identical masses interacting with a single mobile particle moving along the chain~\citep{Mahalingam2023, Mahalingam2023a}.
These studies showed that dissipation is nonlinear, decreases at high particle speeds, and can also be nonmonotonic in particle speed depending on the system parameters.
In particular, if the phonon spectrum is gapped due to the harmonic confinement of the chain masses, the dissipation is exponentially suppressed at low speeds in the absence of thermal vibration.

There are three key takeaways from this work we would like to highlight.
First, by validating the results of Ref.~\citep{Rodin2022} numerically, we confirm the dependence of dissipation on material properties.
Second, by obtaining a time-local equation of motion, we are able to dramatically accelerate simulations of the ionic motion.
This acceleration is useful because ionic motion is a probabilistic process and the quality of numerical experiments improves with simulation time.
Finally, by starting with a simpler 1D system and gradually building up to a more complex 3D configuration, we elucidate the features of the problem that make the time-local treatment possible, paving the way for our future work, where we will include thermal motion of the framework.

In this work, we make use of the techniques developed for 1D systems as the dimensionality constraints are gradually relaxed in order to validate the local-time approximation formalism.
As the first step, we discuss a general model of single-particle motion through a crystal in Sec.~\ref{sec:Model}. 
With the formalism established, readers interested primarily in the final result can skip directly to Eq.~\eqref{eqn:Mobile_EOM_Local} and consult earlier portions of Sec.~\ref{sec:3D_System} for clarifications as needed.
For readers interested in the technical aspects of the formalism, we suggest proceeding in order.
Starting with a 1D chain whose masses can move in three dimensions and can support longitudinal and transverse phonon modes, we use numerical and analytical tools to demonstrate how the system properties determine the energy dissipation of the mobile ion, with the results given in Sec.~\ref{sec:1D_system}.
By exploring the high- and low-speed limits, we establish the range of validity for the local-time approximation.
Next, by focusing on a prototypical 3D system in Sec.~\ref{sec:3D_System}, we show that, unlike the 1D case, where the dissipation diverges at low speeds without harmonic confinement, the drag experienced by slow-moving ions is proportional to their speed.
Finally, we present and validate a time-local formula that connects the framework potential landscape to the dissipation.
We discuss some of the key choices and assumptions used in this paper in Sec.~\ref{sec:Discussion}.
We propose some avenues for extension to thermal systems and conclude in Sec.~\ref{sec:Summary}.

All of our computations were performed using {\scshape julia}~\citep{Bezanson2017}.
The plots were made with the Makie.jl package~\citep{Danisch2021} using a color scheme designed for colorblind readers~\citep{Wong2011}.
The scripts used for computing and plotting can be found at https://github.com/rodin-physics/cubic-lattice-ion-motion.

\section{Model}
\label{sec:Model}

Similar to earlier work,~\citep{Rodin2022, Rodin2022a, Mahalingam2023, Mahalingam2023a} we start with a Lagrangian describing the motion of a mobile particle of mass $M$ through a framework of identical masses $m$:

\begin{equation}
    L = 
     \frac{M}{2}\dot{\mathbf{R}}^T\dot{\mathbf{R}} 
    + \frac{m}{2} \dot{\mathbf{r}}^T \dot{\mathbf{r}}
    -  \mathbf{r}^T \frac{\tensor{V}}{2}\mathbf{r}
    - U(\mathbf{r},\mathbf{R}) \,.
    \label{eqn:Lagrangian}
\end{equation}
Here, $\mathbf{r} = \bigoplus_{j=1} \mathbf{r}_j$ and $\mathbf{R}$ are vectors whose elements are the displacements of all the framework masses from their equilibrium locations and the position of the mobile particle, respectively.
The first two terms in Eq.~\eqref{eqn:Lagrangian} give the kinetic energy of the system, and $\mathbf{r}^T \tensor{V} \mathbf{r} / 2$ is the internal potential energy for the framework using the harmonic approximation.
Finally, $U(\mathbf{r}, \mathbf{R})$ describes the interaction between the framework and mobile particle.

Equation~\eqref{eqn:Lagrangian} makes two important simplifications.
First, it focuses on a single mobile particle.
In real ionic conductors, there are multiple current-carrying ions interacting with each other directly (at small separations) and indirectly (by deforming the framework).
At the same time, the number of framework ions exceeds the number of mobile ions, resulting in a significant separation between the current-carrying ions.
Hence, Eq.~\eqref{eqn:Lagrangian} is applicable when the interaction between mobile ions can be neglected.
The second simplification is the use of the same value $m$ for all framework masses.
Naturally, most materials are composed of multiple atomic types with different masses.
We will demonstrate in Sec.~\ref{sec:3D_System}, however, that this simplification, while making the study much more transparent, does not pose a limitation to our final result, which depends on the density of the system and not the individual masses.

For convenience, we express the coupling matrix in terms of a characteristic spring constant $K$: $\tensor{V} = K \tensor{\Xi}$, to be determined at a later stage.
We can then introduce a characteristic frequency $\Omega = \sqrt{K / m}$, express masses in terms of $m$: $M = \mu \times m$,  and express the evolution time $t$ of the system in terms of the number of periods: $t = \tau \times 2\pi / \Omega$.
Finally, we can express all lengths in terms of the quantum oscillator length associated with $\Omega$, $l = \sqrt{\hbar/m\Omega}$ so that $\mathbf{R} = \boldsymbol{\sigma}l$ and $\mathbf{r} = \boldsymbol{\rho}l$. 
This rescaling turns the Lagrangian into

\begin{equation}
    \mathcal{L} = 
    \frac{\mu}{8\pi^2}\dot{\sigma}^2 
    + \frac{1}{8\pi^2}\dot{\rho}^2
    -  \boldsymbol{\rho}^T \frac{\tensor{\Xi}}{2}\boldsymbol{\rho}
    - \Phi(\boldsymbol{\rho},\boldsymbol{\sigma}) \,,
    \label{eqn:Lagrangian_Dimensionless}
\end{equation}
where we divided both sides of Eq.~\eqref{eqn:Lagrangian} by the characteristic energy scale $\hbar\Omega$, defining $\Phi = U/\hbar\Omega$.
The corresponding equations of motion for the mobile particle and the framework are

\begin{align}
    \frac{\mu}{4\pi^2}\ddot{\boldsymbol{\sigma}} &= - \nabla_{\boldsymbol{\sigma}} \Phi\left(\boldsymbol{\rho},\boldsymbol{\sigma}\right)\,,
    \label{eqn:Mobile_EOM}
    \\
   \frac{1}{4\pi^2} \ddot{\boldsymbol{\rho}} &= -\tensor{\Xi}\boldsymbol{\rho} -\nabla_{\boldsymbol{\rho}}\Phi\left(\boldsymbol{\rho}, \boldsymbol{\sigma}\right)\,,
    \label{eqn:Framework_EOM}
\end{align}
which can be solved in a variety of ways.
In this work, we numerically solve the equations of motion using a Runge-Kutta method.
To explore the dynamics of the framework analytically, however, we also find a formal solution to Eq.~\eqref{eqn:Framework_EOM}.
Dropping the last term yields an eigenvalue equation
\begin{equation}
\frac{1}{4\pi^2} \ddot{\boldsymbol{\rho}} =-\omega_j^2\boldsymbol{\rho}= -\tensor{\Xi}\boldsymbol{\rho} 
    \,,
    \label{eqn:Homogeneous_EOM}
\end{equation}
with normalized eigenvectors $\boldsymbol{\varepsilon}_j$ and corresponding eigenvalues $\omega_j^2$.
The general result for $\boldsymbol{\rho}$ becomes $\tensor{\varepsilon}\boldsymbol{\zeta}(\tau)$, where $\boldsymbol{\zeta}(\tau)$ is a column vector of normal coordinates giving the amplitude of each mode and $\tensor{\varepsilon} = [\boldsymbol{\varepsilon}_1, \boldsymbol{\varepsilon}_2, \dots]$ is a row of column vectors $\boldsymbol{\varepsilon}_j$.
Due to the lattice periodicity, the eigenvector $\varepsilon_j$ is given by $[e^{i\mathbf{L}_1\cdot\mathbf{q}_j}, e^{i\mathbf{L}_2\cdot\mathbf{q}_j},\dots]\otimes \boldsymbol{\eta}_j/\sqrt{N}$.
Here,  $\mathbf{L}_k$ are the coordinates of the $N\rightarrow\infty$ unit cells, $\mathbf{q}_j$ is the momentum associated with the $j$th mode, and $\boldsymbol{\eta}_j$ is a $3n$-dimensional vector with $n$ equal to the number of masses per unit cell.

Using $\boldsymbol{\rho}(\tau) =\tensor{\varepsilon}\boldsymbol{\zeta}(\tau)$, Eq.~\eqref{eqn:Framework_EOM} becomes
\begin{equation}
 \frac{1}{4\pi^2}\ddot{\boldsymbol{\zeta}}  = -\tensor{\omega}^2\boldsymbol{\zeta} -\tensor{\varepsilon}^\dagger\nabla_{\boldsymbol{\rho}}\Phi\left(\boldsymbol{\rho}, \boldsymbol{\sigma}\right) \,,
    \label{eqn:Inhomogeneous_EOM}
\end{equation}
where $\tensor{\omega}^2 = \tensor{\varepsilon}^{\dagger}\tensor{\Xi}\tensor{\varepsilon}$ (using $\tensor{\varepsilon}^{-1} = \tensor{\varepsilon}^\dagger$) is a diagonal matrix of the squared eigenfrequencies.
For each $\zeta_j$, the equation of motion takes the form of a forced harmonic oscillator $\ddot{\zeta}_j = - 4\pi^2\omega_j^2 \zeta_j- 4\pi^2f_j$, which can be solved using the Green's function

\begin{equation}
     G_j(\tau,\tau')= \frac{\sin\left[2\pi\omega_j(\tau - \tau')\right]}{2\pi\omega_j}\Theta(\tau- \tau')\,,
     \label{eqn:Harmonic_Greens_fn}
\end{equation}
to obtain

\begin{align}
    \zeta_j(\tau)
   &= \zeta_j^H(\tau) 
   \nonumber
   \\
   &- 2\pi\int^\tau d\tau'\frac{\sin\left[2\pi\omega_j(\tau - \tau')\right]}{\omega_j}
   \boldsymbol{\varepsilon}^\dagger_j\nabla_{\boldsymbol{\rho}}\Phi[\boldsymbol{\rho}(\tau'),\boldsymbol{\sigma}(\tau')]
   \,,
   \label{eqn:zeta_solution}
\end{align}
where $\zeta_j^H(\tau)$ is the homogeneous solution.
Finally, from Eq.~\eqref{eqn:zeta_solution}, we get

\begin{align}
    \boldsymbol{\rho}(\tau)
   &=  \tensor{\varepsilon}\boldsymbol{\zeta}^H(\tau)
\nonumber
\\
&-2\pi
   \int^\tau d\tau'
   \tensor{\Gamma}(\tau - \tau')
   \nabla_{\boldsymbol{\rho}}\Phi[\boldsymbol{\rho}(\tau'), \boldsymbol{\sigma}(\tau')]
   \,,
   \label{eqn:rho_solution}
   \\
   \tensor{\Gamma}(\tau) & = 
   \sum_j \boldsymbol{\varepsilon}_j \boldsymbol{\varepsilon}_j^\dagger\frac{\sin\left(2\pi\omega_j\tau\right)}{\omega_j}\,.
   \label{eqn:Gamma_mat}
\end{align}
Equation~\eqref{eqn:rho_solution} gives $\boldsymbol{\rho}(\tau)$ as a combination of a homogeneous trajectory, determined by the initial conditions, and a ``memory term'' originating from the framework's interaction with the mobile particle.

As the particle moves through the framework, it loses energy by doing work on the framework masses.
In the absence of thermal motion, the work done up to time $\tau$ is given by~\citep{Mahalingam2023a}

\begin{align}
\Delta(\tau) &= 
\int^{\tau}  d\tau' \left\{-\nabla_{\boldsymbol{\rho}}\Phi\left[\boldsymbol{\rho}(\tau'),\boldsymbol{\sigma}(\tau')\right]\right\}\cdot\dot{\boldsymbol{\rho}}(\tau')
\nonumber
\\
 &=   2\pi^2  \sum_j  \left|\int^{\tau}  d\tau' 
  \nabla_{\boldsymbol{\rho}}\Phi\left[\boldsymbol{\rho}(\tau'),\boldsymbol{\sigma}(\tau')\right] \cdot \boldsymbol{\varepsilon}_j 
  e^{2\pi i\omega_j \tau'}\right|^2 \,.
\label{eqn:Delta}
\end{align}
In the first line, the term inside the curly braces is the force on the framework masses and $\dot{\boldsymbol{\rho}}(\tau')$ their velocity so that the product of these two quantities gives the power exerted by the particle on the framework.
The second line is obtained by substituting the time derivative of the second term in Eq.~\eqref{eqn:rho_solution} for the velocity vector.
In the following two sections, we use Eq.~\eqref{eqn:Delta} as the starting point to study the analytical form of the dissipation.

\section{1D System}
\label{sec:1D_system}

\begin{figure}
    \centering
    \includegraphics[width = \columnwidth]{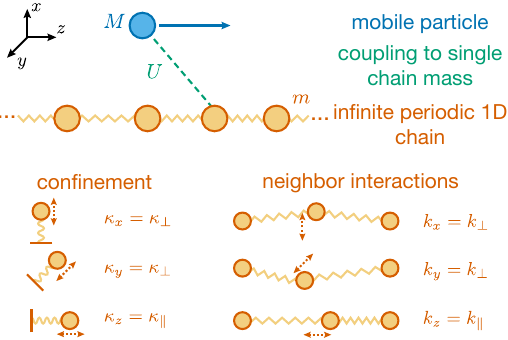}
    \caption{\emph{Schematic of 1D model.} A single mobile particle of mass $M$, confined to motion along the $z$-axis, interacts via a potential $U$ with a single mass which is part of an infinite periodic 1D chain, each of mass $m$. 
    Each chain mass experiences harmonic confinement along all three directions, with spring constants $\kappa_x,\kappa_y,\kappa_z$. 
    The chain masses are also coupled to their nearest neighbors with effective spring constants $k_x,k_y,k_z$. 
    In our simulations, the modes are degenerate, so $\kappa_\parallel=\kappa_\perp$ and $k_\parallel=k_\perp$.}
    \label{fig:1D_Schematic}
\end{figure}

To develop intuition for this form of dissipation, we start with a relatively simple system composed of equally-spaced identical masses arranged in a line, as shown in Fig.~\ref{fig:1D_Schematic}.
At equilibrium, each mass is confined by a harmonic potential in all three orthogonal directions.
Although the chain's masses can move in three dimensions, its vibrational modes are characterized by one-dimensional momentum.
This configuration is distinct from the scenario considered in Refs.~\citep{Mahalingam2023, Mahalingam2023a}, where the masses' motion was restricted to the direction parallel to the chain.
In this section, unless otherwise stated, ``1D chain" refers to a system as shown in Fig.~\ref{fig:1D_Schematic}.

Assuming that the 1D chain of masses runs along the $z$ axis, the force constant in the longitudinal direction is set to $\kappa_\parallel$ and for each of the transverse directions to $\kappa_\perp$.
Additionally, each mass is connected to its immediate neighbors, setting the force constant for the displacement along (perpendicular to) the chain to $k_\parallel$ ($k_\perp$), giving rise to a single longitudinal branch and two branches of transverse modes.
The transverse modes of the chain are distinct from the flexural modes observed in systems like graphene, which have a quadratic dispersion at low momenta.
Here, the dispersion for all three branches, labeled by the index $u$, is given by

\begin{equation}
    \omega_{u,j}^2 = \kappa_u + 4k_u\sin^2\left(\frac{\pi j}{N}\right)\,.
    \label{eqn:omega}
\end{equation}

Following earlier work~\citep{Rodin2022a, Mahalingam2023, Mahalingam2023a}, the interaction between the mobile particle and the framework is given by the sum of pairwise interaction terms, each dependent on the separation between the mobile particle and individual chain masses.
The mass-particle coupling is set to $\Phi(\boldsymbol{\sigma} - \boldsymbol{\rho}_j) = \Phi_0 e^{-|\boldsymbol{\sigma} - \boldsymbol{\rho}_j|^2 / 2\lambda^2}$ for the purposes of demonstration as this potential form is easily tractable analytically and allows for an easy adjustment of the interaction extent and magnitude.

The pairwise interaction and the Gaussian form are major simplifications of the much more complicated $\Phi(\boldsymbol{\rho}, \boldsymbol{\sigma})$ in real materials, where \emph{ab initio} calculations are typically required to obtain this energy.
We discuss the implications of using the simple Gaussian form for more realistic systems in Sec.~\ref{sec:Discussion}.

\begin{figure}
    \centering
    \includegraphics[width = \columnwidth]{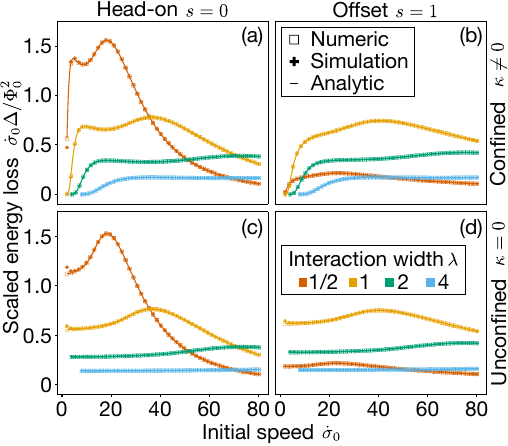}
    \caption{\emph{Dissipation in 1D.}
    After passing a single framework mass, the mobile particle, with initial speed $\sigma_0$, loses energy due to its interaction with that mass.
    The analytic data were obtained from Eq.~\eqref{eqn:Delta_1D}, the numeric results were computed by integrating Eq.~\eqref{eqn:Delta_Integrated} using quadratures, and the simulation was performed as described in the main text using Eqs.~\eqref{eqn:Mobile_EOM}--\eqref{eqn:Framework_EOM}.
    For all panels, $k_\parallel = k_\perp = 99/4$, the chain length was set to 100 with periodic boundary conditions, $\delta\tau = 10^{-4}$, and $\Phi_0 = 0.01$.
    $\boldsymbol{\sigma}_0 = s\hat{y}$ with $s = 0$ for the head-on cases (left column) and $s = 1$ for the offset cases (right column).
    In the top row, $\kappa_\parallel = \kappa_\perp = 1$; in the bottom row, $\kappa_\parallel = \kappa_\perp = 0$.
    For $\kappa \neq 0$, we see dissipation at low speeds vanishes.
    Conversely, for the unconfined chain, $\Delta \propto \dot{\sigma}^{-1}$ at low speeds, leading to flat $\dot{\sigma}\Delta$ curves.
    At high speeds, the dissipation decreases for all configurations.
    }
    \label{fig:1D_Loss}
\end{figure}

To explore the dissipative behavior of this system, we performed several sets of numerical simulations with different system parameters.
For simplicity, we kept the confining potential and the mass coupling isotropic so that $\kappa = \kappa_\perp = \kappa_\parallel$ and $k = k_\perp = k_\parallel$.
With the framework initially at rest, we introduced a particle with mass $\mu = 1$ and speed $\dot{\sigma}$ moving in the positive $z$ direction a distance $7\lambda$ to the left of the interacting mass with one of two impact parameters $\boldsymbol{\sigma}_0 = s\hat{y}$: $s = 0$ or $1$.
We then evolved the system following Eqs.~\eqref{eqn:Mobile_EOM} and \eqref{eqn:Framework_EOM} using the fifth order Runge-Kutta method with time step $\delta \tau = 10^{-4}$, terminating when the particle was more than $7\lambda$ to the right of the interacting mass's initial position.
We computed the loss $\Delta$ by taking the difference between the initial and final kinetic energies of the particle, given by $\mu\dot{\sigma}^2 / 8\pi^2$.
By repeating the simulation for a range of initial speeds, we found the speed-dependent loss for a particular set of framework and interaction parameters,  with results shown in Fig.~\ref{fig:1D_Loss}.

We can gain a better understanding of the nontrivial loss profiles in Fig.~\ref{fig:1D_Loss} by taking advantage of the Gaussian interaction to analyze Eq.~\eqref{eqn:Delta}.
Unfortunately, in its present form, Eq.~\eqref{eqn:Delta} is not tractable because we need $\boldsymbol{\sigma}(\tau)$ and $\boldsymbol{\rho}(\tau)$ which require solving the equations of motion in Eqs.~\eqref{eqn:Mobile_EOM} and \eqref{eqn:Framework_EOM}.
However, if we knew the trajectories, we could directly compute the dissipation.
Making Eq.~\eqref{eqn:Delta} useful requires several simplifying assumptions.
First, we take $\boldsymbol{\rho}$ to be sufficiently small to replace $\boldsymbol{\rho}\rightarrow\boldsymbol{0}$ inside the interaction term of Eq.~\eqref{eqn:Delta}.
Second, we assume the velocity of the particle is constant during the time window of interest, leading to $\boldsymbol{\sigma}(\tau) = \boldsymbol{\sigma}_0 + \tau \dot{\boldsymbol{\sigma}}_0 = s\hat{y} + \dot{\sigma}_0 \tau\hat{z}$.
This approach was previously used in Refs.~\citep{Mahalingam2023, Mahalingam2023a} in the context of ``true" one-dimensional systems where it was shown that this approximation provides reasonable results even when the velocity varies substantially.
Extending the time integration limits to $(-\infty, \infty)$ to obtain the energy lost by the particle during a single ``collision" with a chain mass and taking the time integral yields

\begin{align}
        \Delta &= \frac{2\pi^2}{N}\sum_j
        \left|
         \frac{1}{\dot{\sigma}} 
    \boldsymbol{\eta}_{j}\cdot
    \left[\nabla_\perp + i\hat{z}\frac{2\pi \omega_j }{\dot{\sigma}}\right]
    \tilde{\Phi}_{ 2\pi \omega_j /\dot{\sigma}}\left(\boldsymbol{\sigma}_0\right)
        \right|^2
        \nonumber
        \\
        &= 
   \frac{2\pi^2}{N}  \sum_j  \left| \nabla_\perp\tilde{\Phi}_{2\pi\omega_j/\dot{\sigma}}\left(\boldsymbol{\sigma}_0\right) \cdot \boldsymbol{\eta}_{j}\frac{1}{\dot{\sigma}}\right|^2
  \nonumber
  \\
   &\quad+
   \frac{2\pi^2}{N}  \sum_j  \left|\tilde{\Phi}_{2\pi\omega_j/\dot{\sigma}}\left(\boldsymbol{\sigma}_0\right)\frac{2\pi\omega_j}{\dot{\sigma}^2}\hat{z}\cdot \boldsymbol{\eta}_{j}\right|^2\,.
        \label{eqn:Delta_Integrated}
\end{align}
where $\tilde{\Phi}_{p}(x,y) = \int dz\, e^{ipz}\Phi(x,y,z)$ is the $z$-coordinate Fourier transform of $\Phi(x,y,z)$.
For the Gaussian interaction considered here, $\tilde{\Phi}_{2\pi\omega_j/\dot{\sigma}}(\boldsymbol{\sigma}_0)= \Phi_0\sqrt{2\pi}\lambda \exp\left(-\frac{\sigma_0^2+4\pi^2\omega_j^2\lambda^4/\dot{\sigma}^2}{2\lambda^2}\right)$, setting $\nabla_\perp\rightarrow -\boldsymbol{\sigma}_0/\lambda^2$.
The two terms of Eq.~\eqref{eqn:Delta_Integrated} correspond to two dissipation channels: the first for transverse modes and the second for longitudinal.
The $\boldsymbol{\sigma}_0\cdot\boldsymbol{\eta}_j$ term in the transverse channel arising after the replacement of the gradient operator, combined with the cylindrical symmetry of the system, leads to only the transverse branch with polarization along $\boldsymbol{\sigma}_0$ contributing to dissipation.

Equation~\eqref{eqn:Delta_Integrated} can be evaluated numerically by computing $\boldsymbol{\eta}_j$ and $\omega_j$ from the trivial diagonal dynamical matrix with Eq.~\eqref{eqn:omega} for each direction on the diagonal.
We plot the numerically computed values in Fig.~\ref{fig:1D_Loss} and see an excellent agreement with the simulation results.
The mild deviation seen at small speeds is a consequence of the kinetic energy ($\approx 0.05$ for the slowest speed considered, $\dot{\sigma} = 2$) becoming comparable to $\Phi_0 = 0.01$, violating the constant-speed assumption we employed above.

The simplicity of the system allows us to take the integral in Eq.~\eqref{eqn:Delta_Integrated} analytically.
Because the system contains a single mass per unit cell, the polarization vectors $\boldsymbol{\eta}_j$ contain two 0's and a single 1 corresponding to the polarization direction, leading to

\begin{align}
        \Delta &= 
    \Lambda^2 \Phi_0^2 e^{-\frac{\sigma_0^2}{\lambda^2}}
   \frac{\sigma_0^2}{\lambda^4} 
   \int_0^\pi d\theta \,e^{-\Lambda^2\omega_\perp^2(\theta)}
  \nonumber
  \\
   &+
    \frac{\Lambda^4}{\lambda^2}\Phi_0^2   e^{-\frac{\sigma_0^2}{\lambda^2}}\int_0^\pi d\theta \, \omega^2_\parallel(\theta) e^{-\Lambda^2\omega_\parallel^2(\theta)}
    \nonumber
    \\
    &= 
    \pi \Lambda^2 \Phi_0^2 e^{-\frac{\sigma_0^2}{\lambda^2}}
   \frac{\sigma_0^2}{\lambda^4} 
   e^{-\Lambda^2(\kappa_\perp + 2 k_\perp)}I_0(2\Lambda^2k_\perp)
  \nonumber
  \\
   &+
    \pi\frac{\Lambda^4}{\lambda^2}\Phi_0^2   e^{-\frac{\sigma_0^2}{\lambda^2}}
     e^{-\Lambda^2(\kappa_\parallel + 2 k_\parallel)}
     \nonumber
      \\
      &\times \left[(\kappa_\parallel + 2k_\parallel)I_0\left(2 \Lambda^2k_\parallel\right)-2k_\parallel I_1\left(2\Lambda^2k_\parallel\right)\right]\,,
        \label{eqn:Delta_1D}
\end{align}
where $\Lambda = 2\pi\lambda / \dot{\sigma}$ and $I_n$ are modified Bessel functions of the first kind.
Plotting Eq.~\eqref{eqn:Delta_1D} in Fig.~\ref{fig:1D_Loss}, we see an essentially perfect agreement with the numerical integrals, indicating good convergence.

At high speeds, $\Lambda\rightarrow 0$, resulting in

\begin{equation}
    \Delta_\mathrm{fast} 
    = 
    \pi \frac{\Lambda^2}{\lambda^2} \Phi_0^2 e^{-\frac{\sigma_0^2}{\lambda^2}}\left[
   \frac{\sigma_0^2}{\lambda^2} 
    +
   \Lambda^2 (\kappa_\parallel + 2k_\parallel)\right]\,,
        \label{eqn:Delta_1D_fast}
\end{equation}
showing that the dissipation due to the transverse channel scales as $1 / \dot{\sigma}^2$, while that of the longitudinal channel scales as $1 / \dot{\sigma}^4$.
Naturally, the transverse channel is only possible for nonzero $\sigma_0$ and becomes dominant when $\frac{\sigma_0^2}{\lambda^2} >   \Lambda^2 (\kappa_\parallel + 2k_\parallel)$, corresponding to $\dot{\sigma}  > 2\pi\lambda^2 \sqrt{\kappa_\parallel + 2k_\parallel}/\sigma_0$.
The suppression of $\Delta$ with speed is clearly illustrated by Fig.~\ref{fig:1D_Loss}.

Conversely, at low speeds, $\Lambda\rightarrow \infty$, and

\begin{align}
        \Delta_\mathrm{slow} &= \pi \Phi_0^2 \Lambda e^{-\frac{\sigma_0^2}{\lambda^2}}
    \frac{1}{2\sqrt{\pi}\lambda^2}  
  \nonumber
  \\
   &\times  \Bigg[   
   \frac{\sigma_0^2}{\lambda^2} 
    \frac{e^{-\Lambda^2\kappa_\perp  }}{\sqrt{k_\perp  }}+
      \left(
      \Lambda^2\kappa_\parallel 
      + \frac{1}{2}
      \right)\frac{e^{-\Lambda^2\kappa_\parallel }}{\sqrt{k_\parallel}}\Bigg]\,.
        \label{eqn:Delta_1D_slow}
\end{align}
Equation~\eqref{eqn:Delta_1D_slow} demonstrates a qualitatively different dissipation behavior depending on whether the masses are confined ($\kappa >0$) or not.
For finite $\kappa$, the dissipation becomes exponentially suppressed due to the $\exp\left[-\left(2\pi \lambda \sqrt{\kappa} / \dot{\sigma}\right)^2\right]$ term.
Physically, the time that it takes the particle to pass the chain mass $\sim \lambda / \dot{\sigma}$ is much greater than the period of the slowest vibrational modes $\sim 1/\sqrt{\kappa}$.
Therefore, there is no coupling to the chain modes and, consequently, no dissipation.

Without confinement, $\Delta_\mathrm{slow} \propto \dot{\sigma}^{-1}$.
Physically, at $\kappa = 0$, there are zero-frequency modes, corresponding to translational motion, and the slow particle can couple to them.
Therefore, the particle traveling at a constant speed can be thought as ``dragging" the entire infinitely long chain along with itself.
This divergence is the consequence of the framework's low dimensionality: unlike 3D systems, where the density of states of the zero-frequency modes vanishes, it remains finite in 1D.

For $\kappa = 0$, Eq.~\eqref{eqn:Delta_1D} indicates that the low-speed limit is applicable when $2\Lambda^2k \gg 1$, corresponding to $2\pi^2\lambda^2\omega_\mathrm{max}^2 / \dot{\sigma}^2 \gg 1$, where $\omega_\mathrm{max} = 2\sqrt{k}$ is the maximum frequency of the band.
The $z$-direction Fourier transform of the interaction term also shows that the particle's coupling to a mode $j$ is exponentially suppressed by $\exp\left(-2\pi^2\lambda^2\omega_j^2 / \dot{\sigma}^2\right)$.
Thus, at slow speeds, only low-frequency modes contribute to dissipation so that the actual dispersion of the high-frequency modes is irrelevant, allowing us to linearize the phonon dispersion.
Returning to Eq.~\eqref{eqn:Delta} with the approximation $\boldsymbol{\rho}\rightarrow \boldsymbol{0}$ and integrating over the linearized spectrum, then taking the $(-\infty, \infty)$ time integral yields

\begin{align}
    \Delta 
     &=  \frac{2\pi^2}{N}  \sum_j\sum_{u = \perp,\parallel} \int_{-\infty}^{\infty} 
     d\tau' d\tau
      e^{2\pi i  (\tau'-\tau)2\sqrt{k_u}\frac{\pi j}{N}} 
      \nonumber
      \\
      &\times
      \partial_u \Phi\left[\boldsymbol{\sigma}(\tau')\right]
      \partial_u \Phi\left[\boldsymbol{\sigma}(\tau)\right]
      \nonumber
      \\
      &= 4\pi^3 \sum_{u = \perp,\parallel}   \int_{-\infty}^{\infty} 
     d\tau' d\tau
      \partial_u \Phi\left[\boldsymbol{\sigma}(\tau')\right]
      \partial_u \Phi\left[\boldsymbol{\sigma}(\tau)\right]
      \nonumber
      \\
      &\times
      \delta\left[4\pi^2  (\tau'-\tau)\sqrt{k_u}\right]
      \nonumber
      \\
      &=  \sum_{u = \perp,\parallel} \frac{\pi}{\sqrt{k_u}} \int_{-\infty}^{\infty} 
       d\tau 
      \left[
      \partial_u \Phi\left[\boldsymbol{\sigma}(\tau)\right]
      \right]^2
      \nonumber
      \\
      &=\pi^{3/2}\Phi_0^2\frac{e^{-\frac{\sigma_0^2}{\lambda^2}}}{\lambda\dot{\sigma}}\left(\frac{1}{2\sqrt{k_\parallel}}+\frac{\sigma_0^2}{\lambda^2}\frac{1}{\sqrt{k_\perp}}\right)
      \,,
\label{eqn:Delta_time}
\end{align}
which is identical to Eq.~\eqref{eqn:Delta_1D_slow} with $\kappa = 0$.
The penultimate line in Eq.~\eqref{eqn:Delta_time} shows that, for slow particles, the loss expression becomes local in time.
Based on this result, we will linearize the phonon dispersion for a more complicated 3D system in the following section. 
This approach will allow us to bypass the more complicated approach in Eq.~\eqref{eqn:Delta_Integrated} to directly obtain a local-time result for the dissipation.

\section{3D System}
\label{sec:3D_System}

\begin{figure*}
    \centering
    \includegraphics[width = \textwidth]{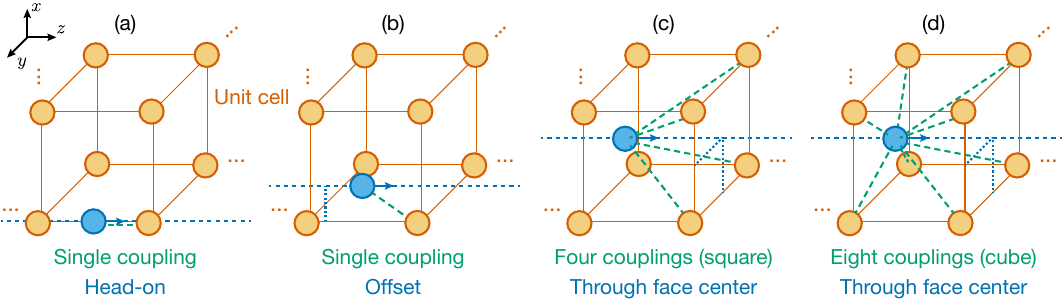}
    \caption{\emph{Schematic of 3D system.}
    We consider four different scenarios for interactions between the mobile ion and 3D the framework, which is a simple cubic lattice in our simulations. In (a), the ion interacts with a single framework mass, and has impact parameter $s=0$. In (b), the ion interacts with a single framework mass, but has impact parameter $\mathbf{s}=\hat{y}$. In (c), the ion interacts with four of the framework masses on a single face of the unit cell perpendicular to the ion's velocity, and passes through the center of that face. In (d), the ion interacts with all eight framework masses in a single unit cell, and passes through the center of two opposite faces.
    }
    \label{fig:3D_Schematic}
\end{figure*}

Before delving into the details of energy loss by a mobile particle in 3D, we explore how the system dimensionality impacts dissipation using a set of simulations similar to the ones performed in Sec.~\ref{sec:1D_system}.
To keep our discussion as transparent as possible, we choose a cubic lattice for the 3D framework.
In the harmonic approximation, the framework masses can be viewed as connected by springs with natural length equal to the framework separation at equilibrium.
Thus, the pairwise potential energy arising from the displacement of two masses is

\begin{align}
    P_{mn} & = \frac{k_{mn}}{2}\left(D_0-|\mathbf{D}_0 - \boldsymbol{\rho}_m + \boldsymbol{\rho}_n|\right)^2
    \nonumber
    \\
    &\approx
    \frac{k_{mn}}{2}\left[\hat{\mathbf{D}}_0\cdot\left(\boldsymbol{\rho}_m - \boldsymbol{\rho}_n\right)\right]^2\,,
    \label{eqn:Potential_Energy}
\end{align}
where $\mathbf{D}_0$ is the vector connecting the equilibrium positions of framework masses $m$ and $n$.
The approximation in the second line holds when $D_0\gg | \boldsymbol{\rho}_m - \boldsymbol{\rho}_n|$.
Assuming that each mass couples only to its first and second nearest neighbors with force constants $k_1$ and $k_2$ respectively, the dynamical matrix becomes~\citep{AshcroftMermin}

\begin{widetext}
\begin{align}
    \mathcal{D}(\mathbf{q}) &= 
    2k_1\begin{pmatrix}
        1 - \cos q_x &0&0
        \\
        0&1 - \cos q_y&0
        \\
       0& 0& 1 - \cos q_z
    \end{pmatrix}
    \nonumber
    \\
    &+2k_2\begin{pmatrix}
      2 - \cos q_x \cos q_y- \cos q_x \cos q_z&\sin q_x \sin q_y & \sin q_x \sin q_z
        \\
         \sin q_y \sin q_x& 2 - \cos q_y \cos q_x- \cos q_y \cos q_z  &  \sin q_y \sin q_z
        \\
        \sin q_z \sin q_x&  \sin q_z \sin q_y&  2 - \cos q_z \cos q_x- \cos q_z \cos q_y 
    \end{pmatrix}\,,
    \label{eqn:D}
\end{align}
\end{widetext}
where $q_{x,y,z} = 2\pi \left[1,2,\dots,N_{x,y,z}\right] / N_{x,y,z}$.

Equation~\eqref{eqn:D} demonstrates why the next-nearest neighbor coupling is required: keeping only $k_1$ decouples the Cartesian coordinates and gives rise to modes with polarization only along the crystal's axes, producing degenerate zero-frequency modes along $\Gamma X$, $\Gamma Y$, and $\Gamma Z$ directions.
Introducing $k_2$ eliminates these nodal lines.
To keep the energy scales similar to the 1D case, we set $k_1 = 15$ and $k_2 = 5$.
The framework masses are not confined, resulting in a phonon spectrum that is not gapped.

\begin{figure}
    \centering
    \includegraphics[width = \columnwidth]{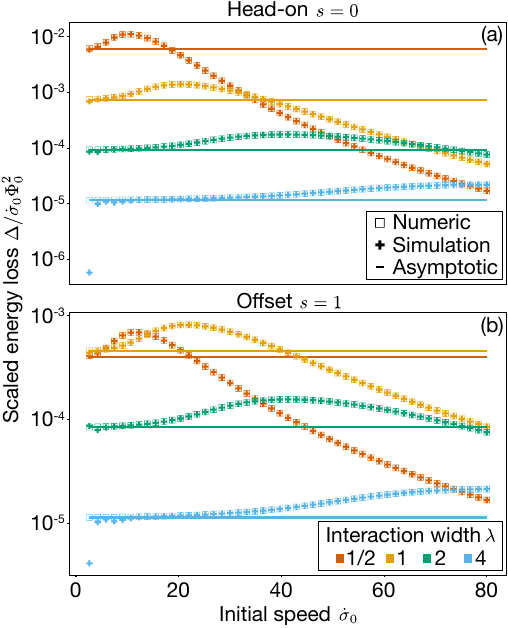}
    \caption{\emph{Dissipation in 3D with a single interaction.}
    When a mobile ion with initial speed $\dot \sigma_0$ passes through a unit cell of a simple cubic lattice, as shown in Fig.~\ref{fig:3D_Schematic}(a) and (b), it loses energy due to interaction with a single mass.
    We computed numeric results by integrating Eq.~\eqref{eqn:Delta_Integrated} using cubatures and obtained the the low-speed asymptotic behavior from Eq.~\eqref{eqn:Delta_3D_Final}, showing that $\Delta \propto \dot{\sigma}$ at low speeds.
    The simulation was performed using Eqs.~\eqref{eqn:Mobile_EOM}--\eqref{eqn:Framework_EOM} with the details given in the main text.
    For both panels, the nearest (next-nearest) neighbor force constant was $k_1 = 15$ ($k_2 = 5$), the system contains $50\times50\times50$ masses with periodic boundary conditions, uses time step $\delta\tau = 10^{-3}$, and has Gaussian interactions with amplitude  $\Phi_0 = 0.01$.
    $\boldsymbol{\sigma}_0 = s\hat{y}$ with $s = 0$ for the head-on collisions and $s = 1$ for the offset configurations.
    The low-speed behavior is captured well by the analytic formula for a range of simulation parameters.
    }
    \label{fig:3D_Loss}
\end{figure}

Identically to the 1D case, the mobile particle interacts with a single framework mass via a Gaussian term and is constrained to move in one dimension, parallel to one of the framework's axes, as shown in Fig.~\ref{fig:3D_Schematic}(a) and (b).
The simulation protocol is the same as in Sec.~\ref{sec:1D_system}, with the particle starting $7\lambda$ to the left of the interacting mass and the Runge-Kutta evolution terminating once the particle reaches the position $7\lambda$ to the right of it.
The results of these simulations are shown in Fig.~\ref{fig:3D_Loss}.

Just as for the 1D case, the loss decays at high speeds.
A crucial difference occurs at low speeds: $\Delta / \dot{\sigma}_0$ approaches a constant, indicating that the dissipation is proportional to the particle speed, instead of diverging as its reciprocal (as is the case for 1D).
This proportionality to speed is reminiscent of drag and is a more intuitive outcome than the divergent behavior for the 1D chain.

We confirm our simulation results with numeric calculations using Eq.~\eqref{eqn:Delta_Integrated}, where we computed the eigenstates using the dynamical matrix in Eq.~\eqref{eqn:D} and integrated over the Brillouin zone using cubatures.
The results of these numeric calculations are shown as open squares in Fig.~\ref{fig:3D_Loss}, and are nearly identical to the simulation results.
They differ only at very small speeds, where the constant-$\dot{\sigma}$ assumption we used to obtain Eq.~\eqref{eqn:Delta_Integrated} is violated.

To explore the low-speed limit, we integrate Eq.~\eqref{eqn:Delta} by parts over both time variables assuming that $\boldsymbol{\rho}\rightarrow \boldsymbol{0}$ and that the mobile particle interacts with a single framework mass:

\begin{align}
    \Delta
     &=   \frac{2\pi^2}{N}  \sum_j \int_{-\infty}^\infty d\tau  d\tau' 
      \nabla \Phi\left[\boldsymbol{\sigma}(\tau')\right]^T \boldsymbol{\eta}_j 
      \nonumber
      \\
      &\times
      \boldsymbol{\eta}_j ^T \nabla \Phi\left[\boldsymbol{\sigma}(\tau)\right]
      e^{2\pi i\omega_j (\tau'-\tau)}
  \nonumber
  \\
  &=
   \frac{1}{N}  \sum_j \frac{1}{2 \omega_j^2}\int_{-\infty}^\infty d\tau  d\tau' 
   \dot{\boldsymbol{\sigma}}(\tau')\cdot
      \mathbf{H}\Phi\left[\boldsymbol{\sigma}(\tau')\right] \boldsymbol{\eta}_j 
      \nonumber
      \\
      &\times
      \boldsymbol{\eta}_j ^T \mathbf{H} \Phi\left[\boldsymbol{\sigma}(\tau)\right]\cdot \dot{\boldsymbol{\sigma}}(\tau)
      e^{2\pi i\omega_j (\tau'-\tau)}
   \,,
\label{eqn:Delta_3D}
\end{align}
where $\mathbf{H}$ denotes the Hessian operator.

To perform the summation over modes, we first write the sum as an integral:
\begin{equation}
     \frac{1}{2N}  \sum_j \frac{\boldsymbol{\eta}_j 
      \boldsymbol{\eta}_j ^T}{ \omega_j^2}
      e^{2\pi i\omega_j \tau}
      =
      \frac{1}{16\pi^3}\int d\mathbf{q}
       \frac{\boldsymbol{\eta}_\mathbf{q}
      \boldsymbol{\eta}_\mathbf{q} ^T}{ \omega_\mathbf{q}^2}
      e^{2\pi i\omega_\mathbf{q} \tau}\,.
      \label{eqn:q_sum_to_integral}
\end{equation}

Next, we write $q_x \rightarrow q\cos\phi\sin\theta$, $q_y \rightarrow q \sin\phi\sin\theta$, and $q_z \rightarrow q\cos\theta$ and linearize the spectrum by expanding the dynamical matrix $\mathcal{D}(\mathbf{q})$ to leading order in $q$, which is $q^2$, so that $\mathcal{D}(\mathbf{q}) \approx q^2 \mathcal{S}(\theta,\phi)$. 
In this limit, the eigenvectors of $\mathcal{D}(\mathbf{q})$ and $\mathcal{S}(\theta,\phi)$ are identical and can be labeled by their branches and directions: $\boldsymbol{\eta}_{u,\theta,\phi}$.
The frequencies $\omega_\mathbf{q}$, on the other hand, are $2\pi q\sqrt{k_{u,\theta,\phi}}$, where $k_{u,\theta,\phi}$ is the eigenvalue corresponding to $\mathcal{S}(\theta,\phi)$'s eigenvector $\boldsymbol{\eta}_{u,\theta,\phi}$, leading to

\begin{align}
     &\frac{1}{2N}  \sum_j \frac{\boldsymbol{\eta}_j 
      \boldsymbol{\eta}_j ^T}{ \omega_j^2}
      e^{2\pi i\omega_j \tau}
      \nonumber
      \\
  =&\sum_u\int_0^\infty dq \int_0^\pi d\theta\sin\theta\oint d\phi \frac{\boldsymbol{\eta}_{u,\theta,\phi} 
      \boldsymbol{\eta}_{u,\theta,\phi}^T}{16\pi^3 k_{u,\theta,\phi}}
      e^{2\pi i\sqrt{k_{u,\theta,\phi}} q\tau}
            \nonumber
      \\
  =&\sum_u  \int_0^\pi d\theta\sin\theta\oint d\phi \frac{\boldsymbol{\eta}_{u,\theta,\phi} 
      \boldsymbol{\eta}_{u,\theta,\phi}^T}{ 16\pi^2 k_{u,\theta,\phi}}
      \delta\left(2\pi \sqrt{k_{u,\theta,\phi}} \tau\right)\,.
      \label{eqn:q_integral}
\end{align}
Combining Eqs.~\eqref{eqn:Delta_3D} and \eqref{eqn:q_integral}, we can write the energy loss in terms of a local-time recoil matrix $\mathcal M$:

\begin{align}
    \Delta
     &=  
  \int_{-\infty}^\infty \underbrace{d\tau  \dot{\boldsymbol{\sigma}}(\tau)}_{d\boldsymbol{\sigma}}
   \cdot
      \,\mathbf{H}\Phi\left[\boldsymbol{\sigma}(\tau)\right] 
    \mathcal{M}\mathbf{H} \Phi\left[\boldsymbol{\sigma}(\tau)\right]\cdot \dot{\boldsymbol{\sigma}}(\tau)
   \,,
   \nonumber
   \\
   \mathcal{M}&=
       \sum_u  \int_0^\pi d\theta\sin\theta\oint d\phi \frac{\boldsymbol{\eta}_{u,\theta,\phi} 
      \boldsymbol{\eta}_{u,\theta,\phi}^T}{ 4\left(2\pi \sqrt{k_{u,\theta,\phi}}\right)^3}\,.
\label{eqn:Delta_3D_Final}
\end{align}
By writing $d\tau\dot{\boldsymbol{\sigma}}\rightarrow d\boldsymbol{\sigma} $, we can identify the remaining portion of the integrand as the dissipative force which is both time-local and proportional to the particle velocity.
For the cubic lattice, numerical integration over $\theta$ and $\phi$ yields an $\mathcal{M}$ that is diagonal with identical entries for each nonzero element, as expected from the system symmetry.

Using this recoil matrix $\mathcal{M}$, we then compute the Hessian of the potential and, assuming a constant speed so that $\boldsymbol{\sigma} = \dot{\boldsymbol{\sigma}}_0\tau+\boldsymbol{\sigma}_0$, take the time integral in Eq.~\eqref{eqn:Delta_3D_Final} numerically to give the low-speed loss.
The results are plotted as straight lines in Fig.~\ref{fig:3D_Loss} for each combination of $\lambda$ and $s$, demonstrating that, at low speeds, Eq.~\eqref{eqn:Delta_3D_Final} agrees very well with the numerical calculations from Eq.~\eqref{eqn:Delta_Integrated}.

\begin{figure}
    \centering
    \includegraphics[width = \columnwidth]{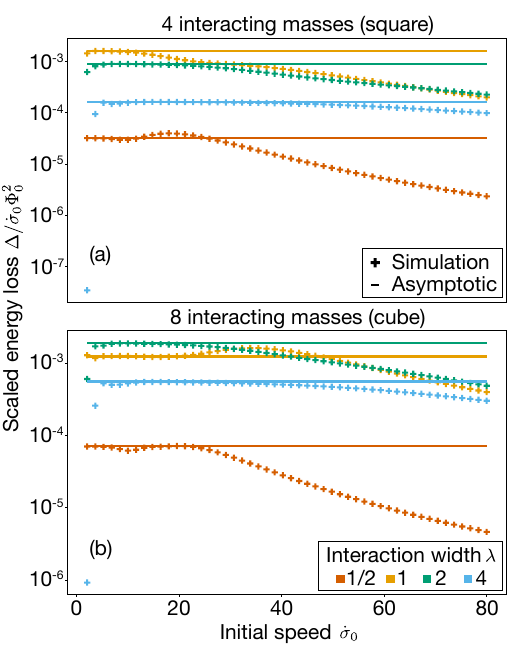}
    \caption{\emph{Dissipation in 3D lattice with multiple interactions.}
    As a mobile particle moves through the middle of a unit cell, it loses energy due to its interaction with (a) four or (b) eight masses in a cubic lattice, as shown in Fig.~\ref{fig:3D_Schematic}(c) and (d).
    The system parameters are the same as in Fig.~\ref{fig:3D_Loss}, but with $\Phi_0$ reduced to $0.0025$.
    The low-speed asymptotic behavior is given by Eq.~\eqref{eqn:Delta_3D_Final}, showing that neglecting the phase difference in the phonon modes of adjacent atoms is valid.
    $\Delta$'s nonmonotonicity with $\lambda$ arises from a combined effect of two opposing phenomena: narrower potentials have a greater second derivative (enhancing dissipation) but the particle spends less time being acted on by the potential (diminishing dissipation).
    }
    \label{fig:3D_Loss_Lattice}
\end{figure}

Our analysis thus far has focused on a rather artificial scenario, where the particle interacts with only a single framework mass.
However, including multiple ion couplings in Eq.~\eqref{eqn:Delta} substantially complicates the expression by introducing a phase prefactor to the vector $\boldsymbol{\eta}$, as discussed in Sec.~\ref{sec:Model}.
Fortunately, our single-interaction results give an important hint that leads to a dramatic simplification.
As we have shown, at low particle speeds, the dissipation is dominated by low-frequency, long-wavelength framework modes.
For these modes, the phase difference between neighboring masses is small, allowing us to neglect it and use Eq.~\eqref{eqn:Delta_3D_Final} with an interaction term $\Phi$ that includes multiple framework masses.

To validate this simplification, we performed two sets of simulations.
For the first set, we initialized the particle normally incident to one face of a cubic unit cell, and set it to interact with the four framework masses defining that face, as shown in Fig.~\ref{fig:3D_Schematic}(c).
As before, the initial position of the particle was $7\lambda$ away from the face of the unit cell and particle was constrained to move in one dimension, passing through the middle of the face.
For the second simulation set, the particle, normally incident to one of the unit cell faces, traversed the entire unit cell while interacting with all eight masses at the corners of the unit cell, see the schematic in Fig.~\ref{fig:3D_Schematic}(d).
In both cases, the simulations terminated once the particle reached a distance of $7\lambda$ from the face containing the interacting masses, at which point its kinetic energy was calculated and subtracted from the initial value.
The results for the energy losses in these two scenarios are shown in Fig.~\ref{fig:3D_Loss_Lattice}.
In addition to the simulation results, we also computed $\Delta$ using Eq.~\eqref{eqn:Delta_3D_Final} for the appropriate form of $\Phi$, shown as horizontal lines in Fig.~\ref{fig:3D_Loss_Lattice}, demonstrating that the asymptotic limit does an excellent job of capturing the behavior of $\Delta$ at low speeds.

Through a series of increasingly realistic interaction scenarios, we have seen that local-time dynamics of the framework in response to its interaction with the particle describe dissipation well.
Although these configurations remain somewhat contrived since the particle moves along a high-symmetry direction, one should keep in mind that their goal is to establish the validity of the local-time approach, not to study realistic particle trajectories.
The fact that the local-time treatment works well regardless whether the particle passes through the potential minimum in the middle of the unit cell or directly through the framework mass demonstrates the robustness of the approach.
Additionally, without including thermal motion of the lattice, which is the subject of our next work, any choice regarding the particle's trajectory is bound to be artificial.

Now that we have confirmed the validity of Eq.~\eqref{eqn:Delta_3D_Final} at low speeds, we can explore the implications of a regime where the force is time-local.
To make use of this simplified formula, we rewrite Eq.~\eqref{eqn:Mobile_EOM} for a small framework deflection by expanding the interaction to leading order in $\boldsymbol{\rho}$ and by using Eq.~\eqref{eqn:rho_solution}, leading to

\begin{widetext}
\begin{align}
    \frac{\mu}{4\pi^2}\ddot{\boldsymbol{\sigma}}(\tau)  
    &=
     - \nabla_{\boldsymbol{\sigma}} \left[\Phi\left[\boldsymbol{0},\boldsymbol{\sigma}(\tau)\right]
     -\nabla_{\boldsymbol{\rho}} \Phi\left[\boldsymbol{0},\boldsymbol{\sigma}(\tau)\right]\cdot 2\pi
   \int^\tau d\tau'
    \sum_j \boldsymbol{\varepsilon}_j \boldsymbol{\varepsilon}_j^\dagger\frac{\sin\left[2\pi\omega_j(\tau-\tau')\right]}{\omega_j} \nabla_{\boldsymbol{\rho}}\Phi\left[\boldsymbol{0},\boldsymbol{\sigma}(\tau')\right]\right]\,.
    \label{eqn:Mobile_EOM_Expanded}
\end{align}
\end{widetext}
Integrating the last term of Eq.~\eqref{eqn:Mobile_EOM_Expanded} over $\tau'$ by parts gives

\begin{align}
   &  \int^\tau d\tau'
   \frac{\sin\left[2\pi\omega_j(\tau-\tau')\right]}{\omega_j}
   \nabla_{\boldsymbol{\rho}}\Phi[\boldsymbol{0}, \boldsymbol{\sigma}(\tau')]
   \nonumber
   \\
   =&
   \frac{1}{2\pi\omega_j^2}
   \nabla_{\boldsymbol{\rho}}\Phi[\boldsymbol{0}, \boldsymbol{\sigma}(\tau)]  
   \nonumber
   \\
   -&
   \int^\tau d\tau'
   \frac{\cos\left[2\pi\omega_j(\tau-\tau')\right]}{2\pi\omega_j^2}
   \nabla_{\boldsymbol{\sigma}}\nabla_{\boldsymbol{\rho}}\Phi[\boldsymbol{0}, \boldsymbol{\sigma}(\tau')]\cdot\dot{\boldsymbol{\sigma}}(\tau')\,.
   \label{eqn:By_parts}
\end{align}
When reinserted back into Eq.~\eqref{eqn:Mobile_EOM_Expanded}, we see that the second term can be identified as giving rise to the dissipative force due to its proportionality to the velocity.
This allows us to replace it by the term from the integrand of Eq.~\eqref{eqn:Delta_3D_Final} multiplying the velocity.
The remaining portion can be combined with the first term in Eq.~\eqref{eqn:Mobile_EOM_Expanded}, resulting in a softened potential so that

\begin{align}
    &\frac{\mu}{4\pi^2}\ddot{\boldsymbol{\sigma}} = 
    \nonumber
    \\
    & - \nabla_{\boldsymbol{\sigma}} \left[\Phi\left(\boldsymbol{\sigma}\right)
     -\nabla_{\boldsymbol{\rho}} \Phi\left(\boldsymbol{0},\boldsymbol{\sigma}\right)\cdot 
    \sum_j \frac{\boldsymbol{\varepsilon}_j \boldsymbol{\varepsilon}_j^\dagger}{\omega_j^2}
   \nabla_{\boldsymbol{\rho}}\Phi(\boldsymbol{0}, \boldsymbol{\sigma})\right]
   \nonumber
   \\
   &-
    \mathbf{H}\Phi(\boldsymbol{\sigma})
     \mathcal{M} \mathbf{H}\Phi(\boldsymbol{\sigma})\cdot \dot{\boldsymbol{\sigma}}\,,
     \label{eqn:Mobile_EOM_Local}
\end{align}
which is the main result of this work.
In this form, we can see that the local-time interaction with the framework gives rise to two phenomena: potential softening and dissipation.
Because both of these terms involve sums $\sim \omega_j^{-2}$, stiff systems (with large $\omega_j$) will be less susceptible to deformation, resulting in a reduced softening and dissipation.
Qualitatively, this result matches our physical intuition.

\begin{figure}
    \centering
    \includegraphics[width = \columnwidth]{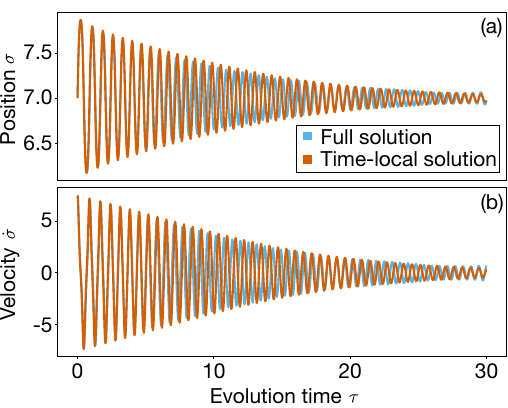}
    \caption{\emph{Local-time dissipation.}
    A particle constrained to move along one of the edges of the unit cell with insufficient energy to escape the local minimum for $\lambda = 1/2$ and $\Phi_0 = 1$.
    The full solution from Eqs.~\eqref{eqn:Mobile_EOM}--\eqref{eqn:Framework_EOM} is obtained using the same procedure as all other simulations.
    The time-local solution is obtained by integrating Eq.~\eqref{eqn:Mobile_EOM_Local} using the fifth order Runge-Kutta method with $\delta\tau = 10^{-3}$.
    The amplitude decay is similar for the two solutions, and the phase difference is related to small displacements of the framework that are neglected by the time-local approach.
    The amplitude increase for the full solution at $\tau\approx 25$ is a consequence of the finite size of the system.
    The calculation times on a single core of a 2020 M1 MacBook Pro for the full solution and the time-local formalism were about 4 hours and about 9 seconds, respectively.
    }
    \label{fig:Local_Time}
\end{figure}

We numerically explore this time-local behavior using the same cubic lattice as in Figs.~\ref{fig:3D_Loss} and \ref{fig:3D_Loss_Lattice}, but extended to include $100\times100\times 100$ masses.
The mobile particle interacts with eight framework masses in a line with equilibrium positions $\alpha(0,0,j)$ for integer $0\leq j \leq 7$ and the lattice constant $\alpha = 2$.
We initialize a mobile particle halfway between the 4th and the 5th masses with speed $\dot{\sigma}$ too small to overcome the energy barrier.
We then compute the trajectory of the particle using the full simulation and Eq.~\eqref{eqn:Mobile_EOM_Local}, then plot the resulting position and velocity evolution in Fig.~\ref{fig:Local_Time}.
Although a more physical trajectory would pass through the middle of the unit cell, similar to the setup in Fig.~\ref{fig:3D_Loss_Lattice}, it would have resulted in a reduced energy barrier and, thus, a much smaller range of possible initial speeds.
Particles moving at these slow speeds would rapidly dissipate their energy after only a few approaches toward the nearby masses.
In order to showcase how non-local time effects can be ignored even when the particle repeatedly interacts with the same masses over an extended period of time, we chose a less-physical setup.

In Fig.~\ref{fig:Local_Time}, we see that the time-local result, both for position and speed, agrees well with the full solution.
The main disparity is the phase difference in the two sets of trajectories, which becomes more pronounced after the particle performs several oscillations around its minimum and the framework masses start to move.
Despite this phase difference, the envelope of the decay agrees very well between the two simulations.
The mild disagreement at larger values of $\tau$ can also be attributed, in part, to the finite size of the system: when we reduced the system size to $80\times80\times 80$, the disagreement increased.
Since these simulations are computationally costly, we chose not to increase the system size beyond $100\times100\times100$.
Naturally, in a real system, thermal motion of the framework will prevent the mobile ion from following such a simple path, so the large-$\tau$ effects will be less relevant.

Up to now, we have been working using dimensionless equations.
To connect our results to real materials, we reintroduce units to Eq.~\eqref{eqn:Mobile_EOM_Local}:

\begin{widetext}
\begin{equation}
    M \ddot{\mathbf{R}} = 
    - \nabla_{\mathbf{R}} \left[U\left(\mathbf{R}\right)
     -  \nabla_{\mathbf{r}}U\left(\boldsymbol{0},\mathbf{R}\right)\cdot 
    \sum_j \frac{1}{m} \frac{\boldsymbol{\varepsilon}_j \boldsymbol{\varepsilon}_j^\dagger}{\Omega_j^2}
  \nabla_{\mathbf{r}}U(\boldsymbol{0}, \mathbf{R})\right]
   -
   \mathbf{H}U(\mathbf{R})
     \left( \frac{2\pi}{m\Omega^3}\mathcal{M}\right) \mathbf{H}U(\mathbf{R})\cdot \dot{\mathbf{R}}\,,
     \label{eqn:Mobile_EOM_Local_Dimensionful}
\end{equation}
\end{widetext}
with

\begin{align}
     \frac{2\pi}{m\Omega^3}\mathcal{M} & =
     \frac{a^3}{m} \frac{1}{(4\pi)^2} \sum_u  \int dS \frac{\boldsymbol{\eta}_{u,\theta,\phi} 
      \boldsymbol{\eta}_{u,\theta,\phi}^T}{ \left(a\Omega\sqrt{k_{u,\theta,\phi}}\right)^3}
      \nonumber
      \\
       & =
     \frac{1}{\rho} \frac{1}{(4\pi)^2} \sum_u  \int dS \frac{\boldsymbol{\eta}_{u,\theta,\phi} 
      \boldsymbol{\eta}_{u,\theta,\phi}^T}{v_{u,\theta,\phi}^3}\,,
      \label{eqn:M_dimensionful}
      \\
      \sum_j \frac{1}{m} \frac{\boldsymbol{\varepsilon}_j \boldsymbol{\varepsilon}_j^\dagger}{\Omega_j^2} & = 
      V^{-1}\sum_j V\frac{\boldsymbol{\varepsilon}_j \boldsymbol{\varepsilon}_j^\dagger}{m\Omega_j^2}
      =
      V^{-1}\,.
      \label{eqn:Softening}
\end{align}
Here the integral is taken over the solid angle, $a$ is the lattice constant, $v_{u,\theta,\phi}$ is the direction-dependent speed of the acoustic branch $u$, and $\rho = m / a^3$ is the framework density.

To make the physical significance of the softening term clearer, we start with the equation of motion for the framework $m\ddot{\mathbf{r}} = -V\mathbf{r} - \nabla_\mathbf{r}U(\mathbf{r}, \mathbf{R})$.
Setting the left-hand side to zero, we obtain the equation which determines $\mathbf{r}$ for which the interaction with the particle is balanced by the elastic forces.
For small framework deformation, we get $\mathbf{r} \approx -V^{-1}\nabla_\mathbf{r}U(\mathbf{0}, \mathbf{R})$.
Thus, the softened potential in Eq.~\eqref{eqn:Mobile_EOM_Local_Dimensionful} is the leading order Taylor expansion of the relaxed configuration.

\section{Discussion}
\label{sec:Discussion}

In the course of our analysis, we have employed a number of simplifications to make the problem tractable.
Having validated these simplifications numerically, let us now address their implications in the context of more general systems.

Our use of the Gaussian interaction was driven primarily by convenience.
At first glance, this form might appear highly artificial because it allows, in principle, for two objects to have zero separation--an unphysical situation for real systems where two ions cannot occupy the same position.
Nevertheless, this choice is less contrived than it might appear as combinations of Gaussian functions are commonly used to describe orbitals in molecular simulations~\citep{Cramer2004}.
By using a single Gaussian function, we treat our ions as spherically symmetric and do not consider ``lobes" in their associated electronic clouds that can arise due to bond formation.
Including multiple Gaussian functions in $\Phi$ makes the problem more complex analytically (e.g., by having Eq.~\eqref{eqn:Delta_Integrated} contain a sum of Fourier transforms), but does not change the approach or the main conclusions.

The cubic geometry was also chosen for its simplicity.
Although real systems generally have a more complex structure, the fact that our approach works well for different dimensionalities supports its robustness.
In fact, the only set of simulations that explicitly reflect the cubic geometry are the ones given in Fig.~\ref{fig:3D_Loss_Lattice}(b): Fig.~\ref{fig:3D_Loss} describes the particle's interaction with a single framework mass, while the square described by the four masses in Fig.~\ref{fig:3D_Loss_Lattice}(a) could belong to a face of a monoclinic unit cell.
Crucially, the local-time formula makes no reference to the lattice geometry as the dissipation matrix $\mathcal{M}$ depends on the speed of sound and material density, while the interaction terms are pairwise and fairly short-range, preventing the mobile particle from ``seeing" the long-range order of the lattice.

Finally, although our study focuses on a single mobile particle, its results are not limited to the case of a single current-carrying ion in an electrolyte system.
Instead, we can think of the model as describing a scenario where the mobile ions are dilute enough so that their interaction can be neglected.
When performing nudged elastic bands calculations, supercells containing a few unit cells in each direction are required to achieve convergence.
Consequently, even separations of a few unit cells between the mobile ions are sufficient to view them as independent.
The role of concerted motion or blocking merits a careful investigation in follow-up work.

\section{Summary}
\label{sec:Summary}

Starting with a simplified 1D setup, in this work we demonstrated that for sufficiently slow-moving particles, the dissipation in solids is dominated by low-frequency, long-wavelength modes.
Using this result, we derived a time-local formula for dissipation in 3D solids, where the dissipative term is proportional to the speed of the moving particle, reminiscent of drag in liquids.
We showed that the drag depends on the curvature of the potential landscape through which the mobile particle moves, as well as the stiffness of the framework lattice.
We validated this newly-derived formula by comparing its results to trajectories computed using the full equations of motion.
Although our formula makes it possible to calculate the trajectory of an individual mobile particle, its chief utility lies in its ability to elucidate the key features of a good ionic electrolyte by doing away with the complexity associated with time nonlocality.
Specifically, because the potential landscape and speed of sound can be computed using \emph{ab initio} methods, our formula can be used as a screening tool for potential electrolyte materials.

A natural extension of this work would investigate the role of thermal fluctuations in the time-local regime.
In this work, we assumed that interaction-induced displacements of the framework are small, allowing us to keep only the first-order correction when calculating the interactions.
In a framework dominated by thermal vibrations, we suspect that neglecting  higher-order contributions to the interaction-driven displacements will be an even better approximation.
Thermal vibrations will effectively reduce the memory of the system, further justifying the local-time approach, but for a more realistic class of systems.    

\acknowledgments

A.R. acknowledges the National Research Foundation, Prime Minister Office, Singapore, under its Medium Sized Centre Programme and the support by Yale-NUS College (through Start-up Grant).
B.\,A.\,O. acknowledges support from the M.\,J. Murdock Charitable Trust.
A.U. was supported by the Ministry of Education, Singapore, under its funding for the Research Centre of Excellence Institute for Functional Intelligence Materials and by the National Research Foundation, Singapore under its AI Singapore Programme (AISG Award No: AISG3-RP-2022-028).

%

\end{document}